\documentclass[12pt,preprint]{emulateapj}

\newcommand{\ltsimeq}{\raisebox{-0.6ex}{$\,\stackrel
        {\raisebox{-.2ex}{$\textstyle <$}}{\sim}\,$}}

\def\msun{{\rm M_{\odot}}}
\def\rsun{{\rm R_{\odot}}}

\def\chandra{{\it Chandra}}

\def\et{{et al.\ }}

\def\fermi{{\it Fermi}}
\def\swift{{\it Swift}}
\def\mon{{Nova Mon 2012}}
\def\cyg{{V407 Cyg}}

\def\H0{{\rm ~km~s^{-1}~Mpc^{-1}}}

\def\msun{M_{\rm \odot}}

\def\et{{et al.}}

\def\deg{^\circ}

\shorttitle{The 7.1~hr period of \mon}
\shortauthors{K.L. Page et al.}

\begin{document}


\title{The 7.1~hour X-ray-UV-NIR period of the $\gamma$-ray classical Nova Monocerotis 2012}


\author{K.L. Page\altaffilmark{1}, J.P
  Osborne\altaffilmark{1}, R.M. Wagner\altaffilmark{2}, A.P. Beardmore\altaffilmark{1}, S.N. Shore\altaffilmark{3}, S. Starrfield\altaffilmark{4}, \& C.E. Woodward\altaffilmark{5}}

\altaffiltext{1}{Department of Physics and Astronomy, University of Leicester, Leicester, LE1 7RH, UK}
\altaffiltext{2}{Department of Astronomy, The Ohio State University, 140 W. 18th Avenue, Columbus, OH, 43210, USA}
\altaffiltext{3}{Dipartimento di Fisica "Enrico Fermi", Universit{\` a} di Pisa, and INFN - Sezione di Pisa, Italy }
\altaffiltext{4}{School of Earth and Space Exploration, P.O. Box 871404, Arizona State University, Tempe, AZ 85287-1404, USA}
\altaffiltext{5}{Minnesota Institute for Astrophysics, U. Minnesota, 116 Church St, SE Minneapolis, MN 55455, USA}
\email{klp5@leicester.ac.uk}

\begin{abstract}

\mon\, is the third $\gamma$-ray
transient identified with a thermonuclear runaway on a white dwarf, that is, a nova event. \swift\, monitoring has revealed the distinct evolution of
the harder and super-soft X-ray spectral components, while \swift\,-UV
and $V$ and $I$-band photometry show a gradual decline with subtle
changes of slope.  During the super-soft emission phase, a coherent 7.1~hr modulation
was found in the soft X-ray, UV, optical and near-IR data, varying in
phase across all wavebands.  Assuming this period to be orbital,
the system has a near-main sequence secondary, with little
appreciable stellar wind. This distinguishes it from the first GeV
nova, \cyg\, where the $\gamma$-rays were proposed to form
through shock-accelerated particles as the ejecta
interacted with the red giant wind. We favor a model in which the
$\gamma$-rays arise from the shock of the ejecta with material close to
the white dwarf in the orbital plane. This suggests that
classical novae may commonly be GeV sources.  We ascribe the orbital modulation to a raised section of an accretion
disk passing through the line of sight, periodically blocking and reflecting much of
the emission. The disk must, therefore, have reformed by day
150 after outburst.

\end{abstract}

\keywords{novae, cataclysmic variables --- stars: individual: Nova Mon 2012  --- ultraviolet: stars --- X-rays: stars}

\section{Introduction}

Nova Monocerotis 2012 is the third GeV nova discovered, with the distinction of being detected in $\gamma$-rays before being discovered at optical wavelengths. The two others detected by the \fermi\,-LAT (Atwood \et\ 2009) are the symbiotic-like nova \cyg\ (Abdo \et\ 2010; Shore \et\ 2011) and Nova Sco 2012 (Cheung \et\ 2012a; Wagner \et\ 2012). For \cyg\, the $\gamma$-rays are believed to be formed either by proton-proton collisions following Fermi-acceleration in shocks formed when the high velocity nova ejecta impact the dense wind from the red giant (RG) companion, or from inverse-Compton emission resulting from the shock-accelerated electrons    
interacting with photospheric photons from the RG (Abdo et al. 2010). 

Novae are due to run-away nuclear
burning on a white dwarf (WD); material is slowly accreted from a binary companion until a critical pressure is reached at the base of the accreted envelope. They occur in binaries with orbital periods of 85~min to several decades (e.g. Munari, Margoni \& Stagni 1990; Bianchini \et\ 2003), the rare examples at wider separations containing an RG with a strong wind, which may under-fill its Roche lobe.

Cheung \et\ (2012b) first announced the discovery of a new $\gamma$-ray source by the \fermi\,-LAT in the Galactic Plane on 2012 June 22. At the time, this area of the sky was too close to the Sun for ground-based observations. However, on 2012 August 9, a new optical nova, \mon, was reported (Fujikawa 2012), which Cheung \et\ (2012c) associated with the \fermi\, source. \mon\, was also detected in the radio before the optical discovery; the radio flux increased in early 2012 September (Chomiuk et al. 2012). O'Brien \et\ (2012) resolved the emission as a double radio source.

High spectral resolution X-ray grating observations were taken by \chandra\, in September and December \citep{ness12b, orioToff12}, 82 and 164 days after \fermi\, discovery. The
first observation showed super-solar abundance, optically-thin
emission from out-flowing plasma with kT~$\sim$~9~keV, while
the second revealed that this plasma had cooled and that an additional
hot WD component had emerged.

Munari (2013) confirmed \mon\, as a neon nova on the basis of strong forbidden Ne {\sc iii - v} emission lines, thus implying a high mass WD.

A 7.1~hr period was first identified in the \swift\, (Gehrels \et\ 2004) data (UV and X-ray; Osborne, Beardmore, \& Page 2013), and subsequently confirmed in the optical (Wagner, Woodward, \& Starrfield 2013). The decline of the super-soft source (SSS) was reported by Page \et\ (2013).
Following on from, and adding to, these data, in this Letter we outline the X-ray/UV/optical/NIR evolution of the nova and show that the near-main sequence secondary implied by an orbital period of 7.1~hr is not inconsistent with a recent model of $\gamma$-ray emission in novae.

Throughout this Letter, day zero has been taken as the time of \fermi\, discovery: midnight on 2012 June 22 (JD 2456100.5). All magnitudes reported are on the Vega system.

\section{Data Collection}

\subsection{Swift: X-ray/UV observations}

\begin{figure*}
\begin{center}
\includegraphics[clip, width=10.0cm,angle=-90]{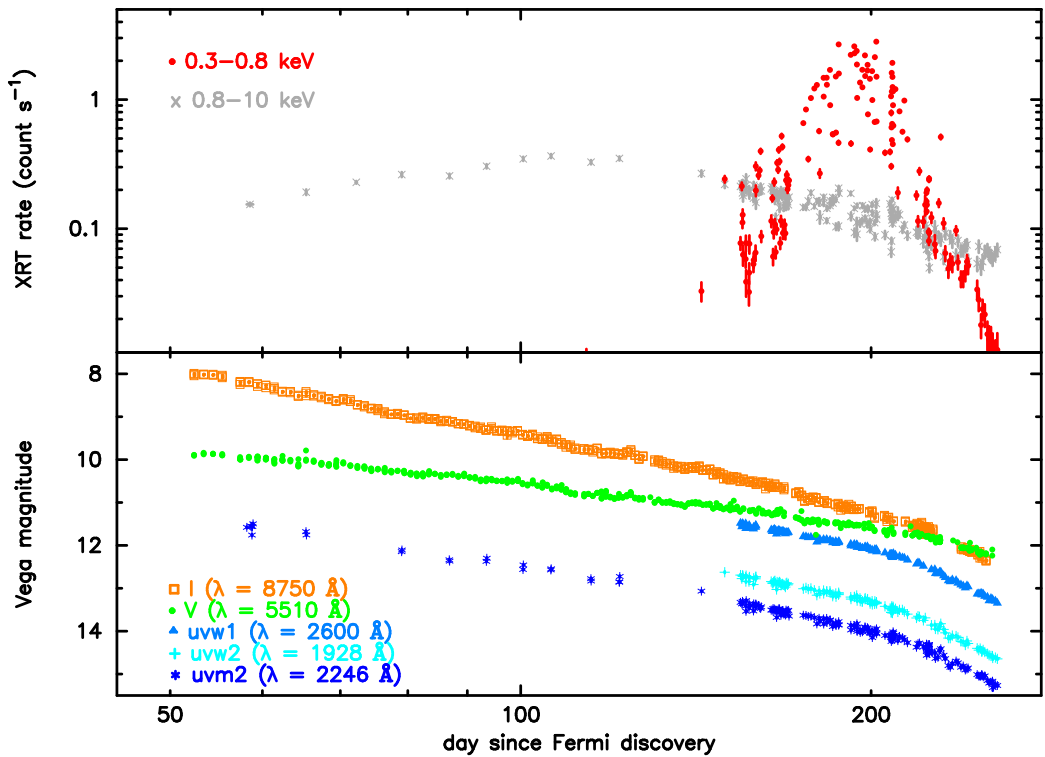}
\caption{Top: XRT light-curves over the soft (0.3--0.8~keV; red) and hard (0.8--10~keV; grey) bands. Before day 140, the soft band count-rate is $<$~0.01 count s$^{-1}$, consistent with that expected from the hard X-ray spectral component. Bottom: MDM and AAVSO $I$-band, AAVSO $V$-band and UVOT data.}
\label{fulllc}
\end{center}
\end{figure*}

After the optical announcement of \mon, weekly \swift\, observations commenced on 2012 August 19. A hard X-ray source was found, fitted by a single-temperature, optically-thin thermal plasma (Nelson \et\ 2012a). 

Shortly after the emergence of the super-soft X-ray emission (2012 November 18; Nelson \et\ 2012b), the cadence of observations was increased, with $\sim$~1--3~ks of data being collected most days. Additionally, on 2013 January 16 a high-cadence campaign started, with snapshots (individual spacecraft pointings) of data taken during each \swift\, orbit for the entire day to confirm an apparent $\sim$~7~hr period in the SSS. On 2013 January 31 and February 1, ten snapshots were obtained simultaneously with ground-based $I$-band data. 

\subsection{MDM: $I$-band photometry}

Differential time-series photometry was obtained in the $I$-band
(725--1025 nm FWHM) on 2013 January 8--13, 30--31, and February 1 UT (days 200--206, 222--223 and 224)
with the 1.3-m McGraw--Hill telescope of the MDM Observatory on Kitt
Peak using the MDM4K CCD imager (1 pixel~=~0\farcs315) windowed to a
field--of--view of 5\farcm4 $\times$ 5\farcm4.  5801 individual
measurements, each with an exposure time of 15~s, were obtained at a
time resolution of 30~s with respect to two anonymous comparison stars
of comparable brightness to the nova. The primary comparison star is
labeled \#117 on the AAVSO finding chart for the nova.  The image
quality was typically 1\farcs2 FWHM, but occasionally reached 2\farcs9 FWHM.  Bias frames and flat--field images of an
illuminated screen in the dome were obtained to facilitate the data
reduction.  In addition, $BVRI$ photometry was obtained
in photometric conditions of the Landolt (1992)
photometric standard star SA98 563, 6\fdg2 from the nova, to calibrate
the $I$-magnitudes of the two comparison stars used to place the nova
magnitude on an absolute flux scale. 

\subsection{AAVSO: $I$ \& $V$ photometry}

$V$ and $I$-band photometry was obtained from the American Association of Variable Star Observers
(AAVSO; see also Hambsch, Krajci, \& Banerjee 2013). The majority of the AAVSO observers provided consistent measurements; values from those which did not were excluded. These data, as well as our $I$-band photometry obtained after day 200, are plotted in Fig.~\ref{fulllc}.

\section{Data Analysis}

\subsection{Swift}


\swift\, data were processed using HEASoft 6.12, following standard procedures. All X-ray Telescope (XRT; Burrows \et\ 2005) data were collected in Photon Counting mode; we used event grades 0--12. When the source X-ray count-rate was $\ltsimeq$~0.6~count s$^{-1}$, a circle of 15--20~pixels (1~pixel~=~2.36\arcsec) was used to extract the flux. Otherwise an annulus was used to avoid pile-up in the inner region of the point spread function, excluding three to seven pixels radius, depending on brightness.
UV/Optical Telescope (UVOT; Roming \et\ 2005) magnitudes were calculated using a 5\arcsec\ source region. For both instruments, larger circular regions close to the source were used for background estimation.
There was no detection of \mon\, by the \swift\, Burst Alert Telescope (15--350~keV; Barthelmy \et\ 2005) in 2012 June (H.A. Krimm, priv. comm.).

\subsubsection{XRT}

\swift\,-XRT data were extracted over 0.3--10~keV for the full band light-curve, with the hard and soft bands being defined as 0.3--0.8 and 0.8--10~keV.  Fig.~\ref{fulllc} (top) shows the soft and hard X-ray  light-curves, with a single bin per snapshot. Before about day 140, the spectrum can be modelled by an absorbed, single-temperature, optically-thin component (Nelson \et\ 2012a).  The absorbing column was initially high, declining by $\sim$~day 80 to around the interstellar value estimated by Shore \et\ (2013). At later times, an additional, much softer emission component was required to model the emergence of the SSS. This had negligible counts above 0.8~keV, hence the definition of the bands above. This soft component was variable by up to a factor of ten within a day. 


\subsubsection{UVOT}

The \swift\,-UVOT  initially collected data in the $uvm2$ filter (central wavelength 2246 \AA; FWHM 498 \AA). Once the super-soft emission emerged, most observations obtained data in all three UV filters: $uvw2$ (1928 \AA; FWHM 657 \AA), $uvm2$ and $uvw1$ (2600 \AA; FWHM 693\AA), see Fig.~\ref{fulllc}.

\subsection{MDM Photometry}
%


The images were reduced and analyzed using IRAF 2.16.  Each individual image of the nova was
bias--subtracted and then corrected for pixel--to--pixel response variations using a single median flat-field image constructed from the
set of flat-field images. The brightness of the nova
(v) and the two comparison stars (c and k) was measured on each image
using circular aperture photometry with annular background
subtraction.  The differential light-curves were constructed by
differencing the observed instrumental magnitudes $(v-c)$ and $(c-k)$.
The photometric precision of the differential nova $(v-c)$ photometry
was $0.001-0.002$ magnitude based on the $(c-k)$ measurements.
Finally, the nova photometry was placed on an absolute scale by the
calibration of the primary comparison star c for which we measured $I$~=~11.18~$\pm$~0.04~mag based on
the standard star observation.

\section{Results}

\subsection{Spectral Evolution}

\begin{figure}
\begin{center}
\includegraphics[clip, width=3.8cm,angle=-90]{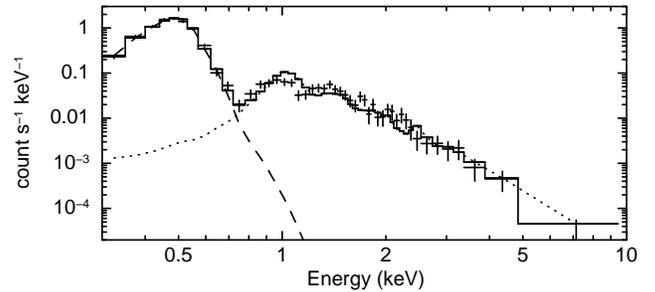}
\caption{X-ray spectrum covering days 175--181 after the \fermi\, discovery. The model comprises a WD atmosphere (kT~$\sim$~65~eV; dashed line) and a single-temperature optically-thin plasma component (kT~$\sim$~1.2~keV; dotted line).}
\label{spec}
\end{center}
\end{figure}

As demonstrated by  the soft and hard X-ray light-curves in Fig.~\ref{fulllc}, there is clearly more than one emitting  component after day 150. 
The spectrum was modelled with a combination of a plane-parallel,
static, non-local thermal equilibrium atmosphere
component\footnote{We used grid model 003 from http://astro.uni-tuebingen.de/\raisebox{.2em}{\tiny
$\sim$}rauch/TMAF/flux\_HHeCNONeMgSiS\_gen.html} (Rauch 2003; Rauch \et\ 2010) to parameterize the soft, WD emission, and a single-temperature optically-thin thermal plasma component at higher energies, both absorbed by a freely-varying column (see Fig.~\ref{spec}). 
The bolometric luminosity estimated from these WD atmosphere fits, for a distance of 3.6~kpc (Shore \et\ 2013), was around 0.1--1 times the Eddington luminosity for a 1.2$\msun$ star. Although a full spectral analysis is beyond the scope of this Letter, no substantial spectral evolution in these two emission components, apart from the overall intensities, was seen.

At longer wavelengths, the nova faded smoothly and at an increasing rate; over extended intervals magnitudes are proportional to log(time), i.e. $f~\propto~(t/1~day)^{\alpha}$. The light-curves break at day~$\sim$~150 and more steeply at day~$\sim$~220. The $uwm2$ and $I$-bands decline similarly after day~150, while the $V$-band decline is always markedly slower; thus, $uvm2-V$ gets redder, while $V-I$ gets bluer. Between days 60--150, $\alpha$~$\sim$~1.6, 1.1, 2.1 in the $uvm2$, $V$ and $I$-bands respectively. Then until day 220, the $uvm2$ and $I$-bands showed $\alpha$~$\sim$~2.8, with $\alpha_V$~$\sim$~1.5. After this time, the decline indices doubled. We speculate that the slow decline in the $V$-band may be due to persistent line emission (e.g. O and Ne; Shore \et\ 2013).

\subsection{Timing}

\begin{figure}
\begin{center}
\includegraphics[clip, width=2.88cm,angle=-90]{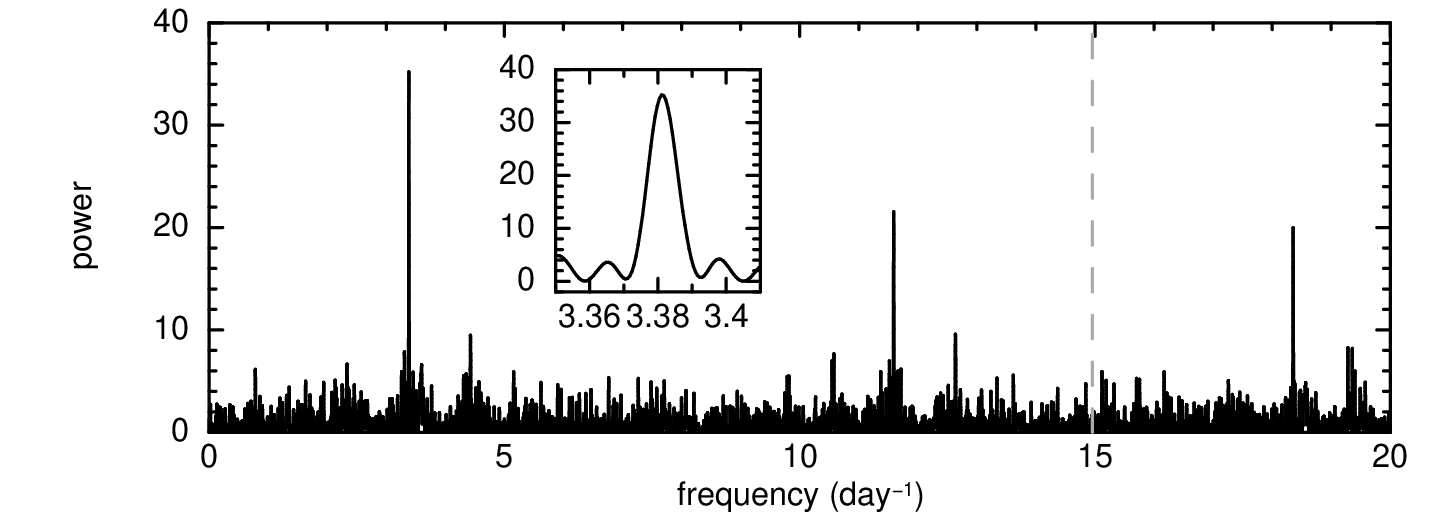}
\includegraphics[clip, width=5.5cm,angle=-90]{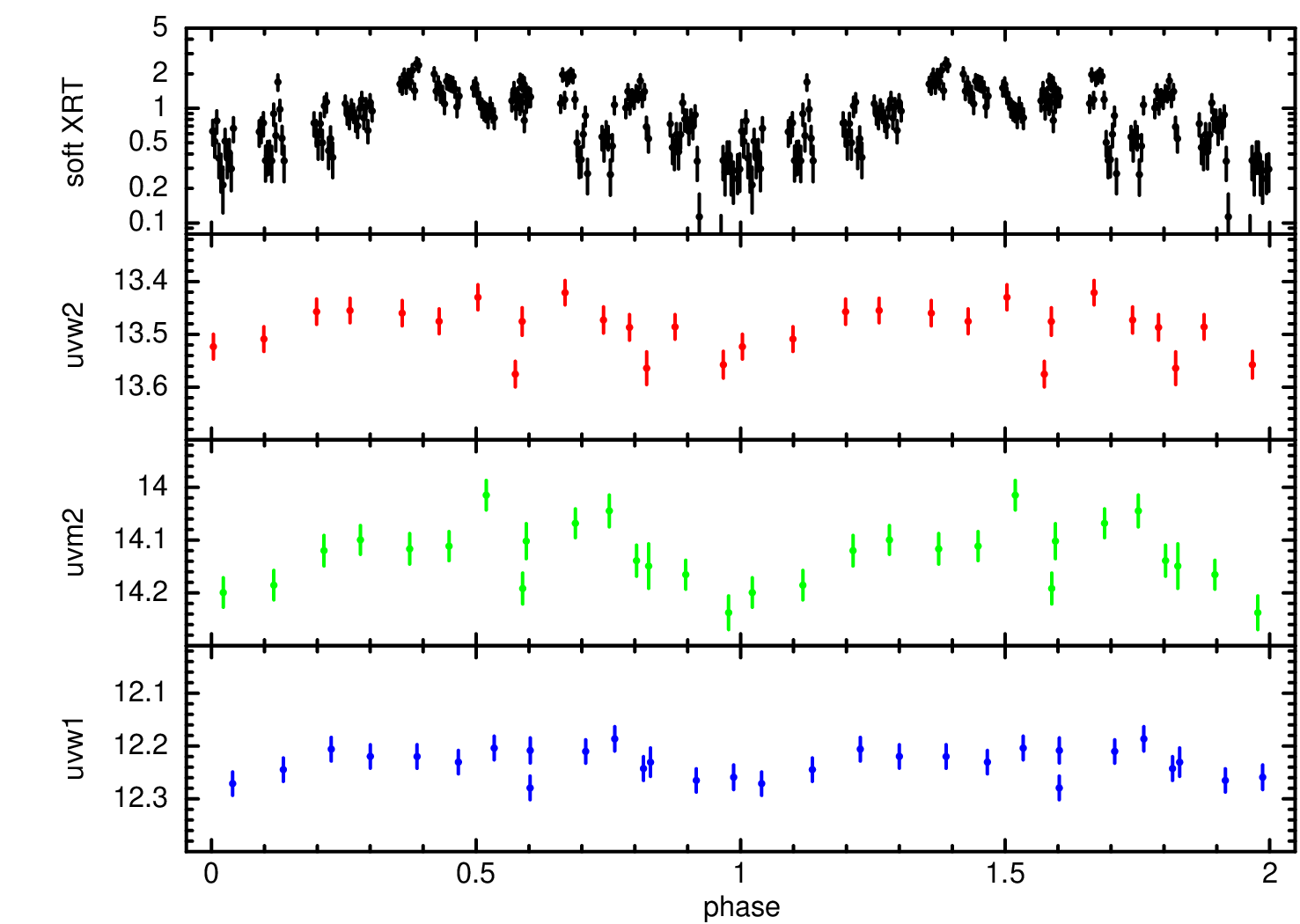}
\includegraphics[clip, width=2.1cm,angle=-90]{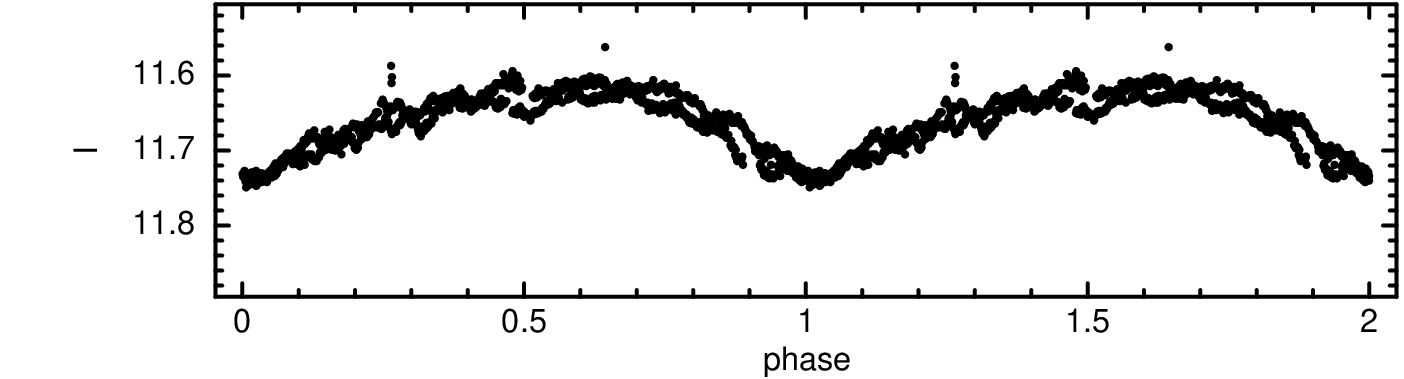}
\caption{Top: Lomb-Scargle periodogram of detrended UVOT $uvm2$ data after day 150. The strong spike at 3.3817 day$^{-1}$ (see inset) shows the period of \mon\,. The dashed line indicates the frequency of \swift's orbit; the symmetrically-placed peaks either side are aliases of the nova and \swift\, periods. Middle: High-cadence \swift\ data (days 208--209) folded on a period of 0.29571 days. 100 s bins were used for the soft X-ray light-curve and the UV data have one bin per snapshot. Bottom: $I$-band data from days 223--224, folded on the same period.}
\label{foldlc}
\end{center}
\end{figure}

A Lomb-Scargle periodogram of the \swift\, data after day~150 showed evidence for a 7.1~hr period in both the UVOT and XRT light-curves.
The top panel of Fig.~\ref{foldlc} shows the periodogram of the UVOT $uvm2$ data taken after the onset of super-soft emission (days~150--257), following detrending by subtraction of the best-fit third-order polynomial. (The earlier data are
too sparse to help constrain a period of a few hours.)

A folding search technique was used to measure the period (e.g., Leahy 1987; Larsson 1996),
whereby $\chi^2$ is calculated as a function of trial period when the
heliocentrically-corrected data are folded at each period (into eight phase
bins). Fitting the resulting distribution of $\chi^2$ near its maximum
gave a period of 0.29571~$\pm$~0.00010~day, where the error is at 90\% confidence and was estimated by applying the same technique to
Monte Carlo simulations of the data sampled at the same
times as the observations, with appropriate Gaussian noise added.

The resulting  UTC ephemeris derived from the UVOT $uvm2$ data is 
\begin{equation} 
T_{min} = HJD~2456300.6684 (50) + 0.29571 (10)~E 
\end{equation} 
where E is the cycle count and the numbers in parentheses indicate the 90\% errors, with the uncertainty on the epoch determined by
estimating the error in the centroid of a Gaussian fit to the minimum
seen in the phase-folded light-curve. Consistent results were obtained from the other UVOT filters and NIR photometry.

Fig.~\ref{foldlc} (middle) plots the high-cadence \swift\, data
obtained throughout 2013 January 16, folded according to Equation~1. The modulation is clearly seen in the
soft X-rays (0.3--0.8~keV) and all three UV filters. Fig.~\ref{foldlc} (bottom) shows the folded $I$-band data from days 223--224. The 90\% upper limit on the modulation in the hard
(0.8--10~keV) X-ray data is 5\%. Fitting the XRT spectrum from January 16 with a combination of an atmosphere model and optically-thin thermal emission,
we found that the optically thin component provides $\sim$~10\% of the counts in this soft band. Within this band there is no discernible spectral variation as a function of the 7.1~hr phase.


\begin{figure}
\begin{center}
\includegraphics[clip, width=6cm,angle=-90]{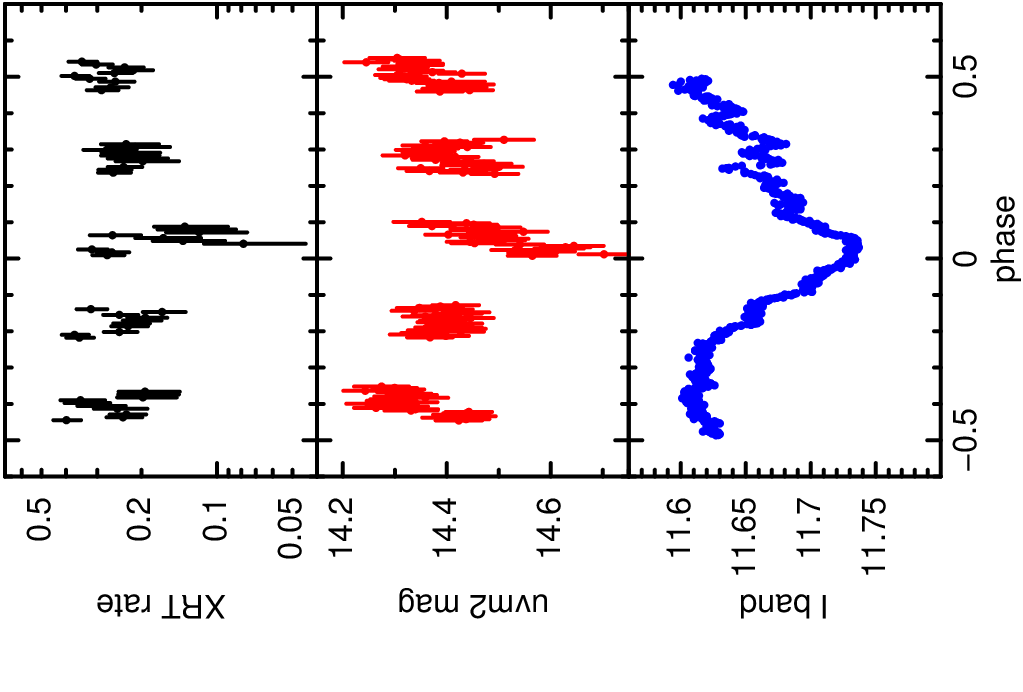}\hspace{0.1cm}
\includegraphics[clip, width=6cm,angle=-90]{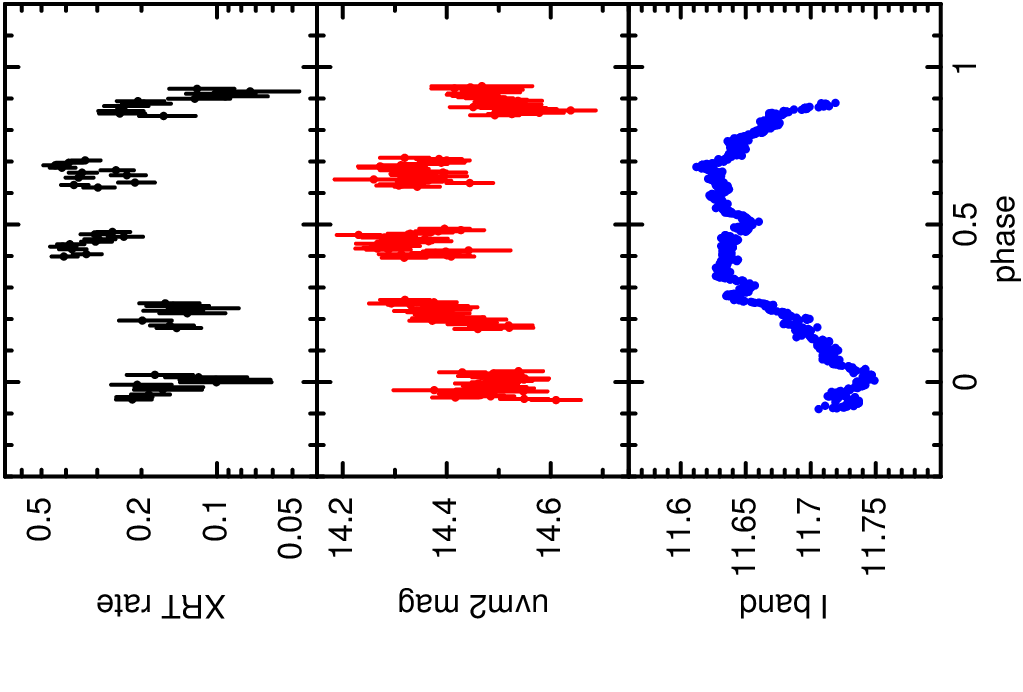}
\caption{Simultaneous \swift\,-X-ray (0.3--10 keV) and I-band photometry, plotted against phase. The X-ray points are binned every 200~s, while the UV points are every 100s.}
\label{simlc}
\end{center}
\end{figure}

Fig.~\ref{simlc} shows the data collected simultaneously by \swift\,-XRT, -UVOT and MDM. Short-term variations are seen across all three bands.

\begin{figure}
\begin{center}
\includegraphics[clip, width=5cm,angle=-90]{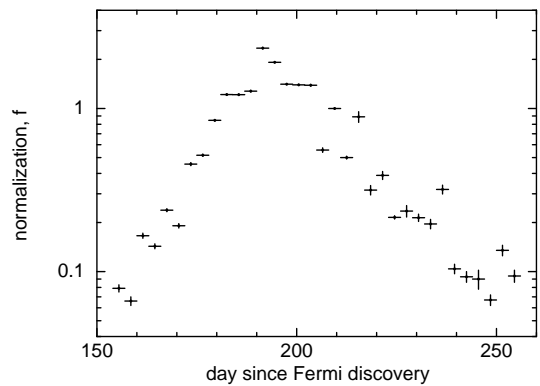}
\caption{Normalization of a constant modulation model fit to 3-day folds of the XRT data.}
\label{mult}
\end{center}
\end{figure}

The large amplitude variability of the super-soft X-ray emission (Fig.~\ref{fulllc}) superficially appears random; however, it can be decomposed into a slower intensity evolution and the observational sampling of the 7.1~hr periodic modulation.  Noting that the amplitude range, for much of the super-soft phase, is around a factor of ten, we constructed a count-rate model as a function of phase, of the form $C = f\cdot(\rm{A}+\rm{B~sin}~\phi)$ count s$^{-1}$. The coefficients A and B were determined from an early section of the data (A/B~=~1.71). The normalization parameter, $f$, was measured by fits to 3-day sequences of observations, a compromise between desired time resolution and sufficient phase coverage (Fig.~\ref{mult}). The point-to-point absolute variability in this plot averages $\sim$40\%, much less than the $\sim$75\% of Fig.~\ref{fulllc}. Much of the fast variation in Fig.~\ref{fulllc} is due to the periodic modulation; removing this reveals the more steadily rising and declining super-soft source.  
The large-amplitude variability seen in the early SSS emission of RS~Oph (Osborne \et\ 2011) is not present in this system.

\section{Discussion}

\swift\, monitoring of the GeV ONe \mon\, revealed a slowly-evolving, hard X-ray-emitting, optically-thin plasma and a much more variable super-soft X-ray component, which was first seen on day 150 and which faded to 2\% of its peak rate by day 250. 
As shown by Schwarz \et\ (2011), the late turn-on time of the SSS suggests an explosion which was unusually energetic  within that sample, with an ejecta mass consistent with that reported by Shore \et\ (2013). 
The time of the start of the decline of the SSS, around day 200, is consistent with the H$\alpha$ emission-line velocity\footnote{The true ejecta velocity may be larger.} of 2400 km~s$^{-1}$ FWHM (Shore et al. 2013), according to the relationship of Greiner, Orio, \& Schartel (2003). The turn-off time suggests a two magnitude decline time from maximum of t$_{\rm 2}$~$\ltsimeq$~10~day (Schwarz \et\ 2011), placing \mon\, in the very-fast class.
The speed-class, together with the SSS turn-off time, are commensurate with the UV line-ratio-determined nuclear turn-off times measured by Vanlandingham \et\ (2001) for ONeMg novae. The WD temperature and t$_2$ time are consistent with the correlation of Orio (2012), while the WD temperature requires a WD mass of $\sim$~1.25$\msun$ (Starrfield \et\ 2012), consistent with its ONe composition. In summary, as an ONe nova, \mon\, appears to be a typical member of its class. We now consider the implications of the main result of this Letter.


We found a coherent 7.1~hr periodic modulation in the soft
X-ray, UV, optical and NIR photometry of \mon\,
during the epoch when the SSS is visible. This modulation
is in phase over all wavebands,  with the soft
X-rays originating from the hot WD.
The
7.1~hr period is almost certainly the orbital period of the
system. We ascribe the modulation to the effects of an orbiting disk rim bulge viewed at moderately high inclination. Novae occur in binary systems with a wide range of
orbital periods, with most below one day (Warner
2008). On-going accretion requires that the secondary star fills its
Roche lobe which, for orbital periods $\ltsimeq$8~hr, dictates that
it be close to the main sequence. In this case the secondary star mass is M$_2$~$\simeq$~0.8~$\msun$, and it will have little appreciable wind. In contrast, symbiotic novae have orbital periods of order years, and giant secondaries with dense winds. The orbital
period of \mon\, makes it distinct from the GeV nova \cyg\, which has an orbital period of 40--50~yr
(Munari \et\ 1990; ; Shore \et\ 2011), a Mira RG secondary and, thus, a dense wind
which acts as a target in the shock generation giving rise to
the $\gamma$-ray emission. 


Shock acceleration of particles to TeV energies can account for the lower shock
velocity determined from X-ray temperature measurements than that
measured by emission line widths in the IR, as was the case in
the symbiotic nova RS~Oph (Tatischeff \& Hernanz
2007). Energy loss from the shock is due to very high energy
particles dominating the photon luminosity by a factor $\sim$~200, thus
motivating the $\gamma$-ray emission model (Abdo \et\ 2010). The inverse-Compton and proton-proton GeV $\gamma$-ray mission models of \cyg\, were
considered in more detail by Martin \& Dubus (2013), who showed
that a better fit to the \fermi\,-LAT data was
achieved by a model in which a density enhancement around the WD gives rise to a dominant inverse-Compton component,
as opposed to the original RG wind model. 

Density enhancement in the orbital plane is a natural consequence of
the presence of the WD, which gives rise to an accretion disk with a high angular momentum outflow in the plane, and to
an accretion wake in the case of a significant secondary wind. \mon\, does not have such a
wind, so we must look to the accretion disk and related structures
for the material against which the ejecta shock. Ejecta structure
observed at high inclination is hinted at by the detection of a double
radio source (O'Brien \et\ 2012). The presence of such structure
is also consistent with the UV and optical emission line analysis of
Shore \et\ (2013), who show that the horned line profiles are
consistent with inclinations around 50--70$\deg$\footnote{After our paper was submitted, Ribeiro \et\ (2013) estimated $i$~=~82~$\pm$~6$\deg$}. 
Highly
non-spherical nova ejecta have been derived by Walder, Folini and
Shore (2008) and Drake \et\ (2009) for the symbiotic nova RS~Oph, formed by an orbital plane density enhancement. 
The same mechanism may be at play in \mon\,. 

We presume that
the $\gamma$-ray emission of \mon\, is due to shock-accelerated electrons, generated by the nova ejecta impacting the
accretion disk or related structure, and boosting the nova light as in the
Martin \& Dubus (2013) model for V407~Cyg. In their favored model, the secondary star wind shock was not a significant source of $\gamma$-rays. Similarly, Shore \et\ (2013) conclude that internal shocks in the ejecta, rather than a wind shock, are the cause of the GeV emission. The much smaller system dimensions of \mon\, compared to \cyg\ suggest a shorter GeV emission episode; however a more complete test of this hypothesis, also making use of the harder X-ray emission, is beyond the scope of this Letter.

Our discovery of a 7.1~hr
modulation shows that the accretion disk has reformed by day 150, and that the
binary system is seen at a moderately high inclination. X-ray
modulation, such as seen here, occurs due to the passage though the
line of sight of a raised sector of the accretion disk rim where the
accretion stream from the secondary impacts the disk, possibly augmented by reflection around phase 0.5. Such a disk was seen to be reforming between day 23--35 in the recurrent nova U~Sco (Ness \et\ 2012a).
Treating the
emitter as a point source, the lack of evident eclipses by the
secondary star (R~$\simeq$~0.8$\rsun$) restricts the system inclination to $i$~$\ltsimeq$~70$\deg$. The fact that the modulation is in phase from the X-ray to
the $I$-band suggests that the modulated emission in all these bands originates
interior to the disk rim. Residual nuclear burning on the WD,
presumably also fed with fresh hydrogen-rich fuel by the accretion disk, is
the cause of the observed super-soft X-rays, while the emission at longer wavelengths
likely arises from the intensely illuminated disk itself (Suleimanov, Meyer, \& Meyer-Hofmeister 1999; 2003).
The broad shape of the X-ray modulation, with the dip covering around half the
orbital cycle, implies an extended absorbing or emission region. An optically-thin electron-scattering
cloud, potentially originating in gas stripped from the accretion disk
surface, may be the cause of an extended emission region. Beardmore \et\ (2012) made a similar inference for the super-soft nova HV~Cet which
also exhibited in-phase X-ray, UV and optical orbital modulation.

\section{Acknowledgments}

We thank the \swift\, PI and MOC, and the staff of the MDM Observatory for their support of these observations.
We acknowledge the
observations from the AAVSO International Database contributed
by observers worldwide. 
The \swift\, project at the University of Leicester is supported by the UK Space Agency. SS acknowledges partial support from NASA and NSF grants to ASU.


\end{document}